\documentclass[%
 aip,
 amsmath,amssymb,
 reprint,%
]{revtex4-1}

\usepackage{graphicx}
\usepackage{dcolumn}
\usepackage{bm}

\usepackage[utf8]{inputenc}
\usepackage[T1]{fontenc}
\usepackage{mathptmx}
\usepackage{textcomp}
\usepackage{color}
\begin{document}

\preprint{AIP/123-QED}

\title{Frequency comb generation threshold in $\chi^{(2)}$ optical microresonators}

\author{Jan Szabados}
\affiliation{ 
Laboratory for Optical Systems, Department for Microsystems Engineering - IMTEK, University of Freiburg, Georges-K\"ohler-Allee 102, 79110 Freiburg, Germany
}%
\author{Boris Sturman}%
\affiliation{ 
Institute of Automation and Electrometry, Russian Academy of Sciences, Koptyug Avenue 1, 630090 Novosibirsk, Russia}%
\author{Ingo Breunig}
\email{optsys@ipm.fraunhofer.de.}
\affiliation{ 
Laboratory for Optical Systems, Department for Microsystems Engineering - IMTEK, University of Freiburg, Georges-K\"ohler-Allee 102, 79110 Freiburg, Germany
}%
\affiliation{ 
Fraunhofer Institute for Physical Measurement Techniques IPM, Heidenhofstra\ss e 8, 79110 Freiburg, Germany
}%

\date{\today}

\begin{abstract}
We investigate the threshold of $\chi^{(2)}$ frequency comb generation in lithium niobate whispering gallery microresonators theoretically and experimentally. When generating a frequency comb via second-harmonic generation, the threshold for the onset of cascaded second-order processes leading to a comb is found to be approximately 85 \textmu W. The second-harmonic generation efficiency up to this value is in excellent agreement with a previously known theoretical framework. This framework is extended here, showing that the onset of cascaded $\chi^{(2)}$ processes and the maximum of the second-harmonic generation efficiency coincide. Furthermore, we observe that the frequency distance between the comb lines is a function of the pump power. It changes from 4 free spectral ranges at the oscillation threshold to 1 free spectral range at 590~\textmu W.
\end{abstract}

\maketitle

\section{\label{sec:level1}Introduction}

\begin{figure}
	\centering
	\includegraphics{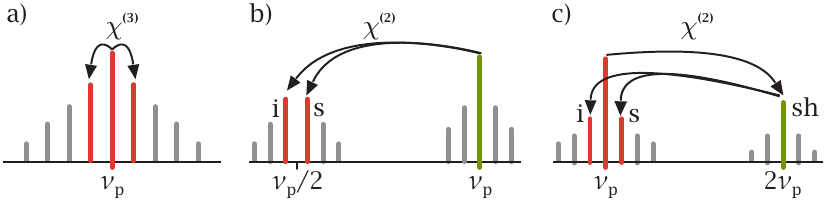}
	\caption{Different starting processes for frequency comb generation. \textbf{a)} Using third-order $\chi^{(3)}$ nonlinearities, comb lines are generated around the pump frequency $\nu_\mathrm{p}$. In \textbf{b)}, $\nu_\mathrm{p}$ drives a $\chi^{(2)}$ optical parametric oscillation (OPO) process, resulting in signal (s) and idler (i) light around half the frequency. Subsequent $\chi^{(2)}$ nonlinear-optical processes such as second-harmonic and sum-frequency generation lead to the build-up of combs around both frequencies. In \textbf{c)}, the initial step for comb generation is second-harmonic generation (sh). When the power of the second-harmonic light overcomes the threshold for OPO, a cascade of $\chi^{(2)}$ processes leads to the build-up of combs around both frequencies.}
	\label{fig1}
\end{figure}

Optical frequency combs have attracted considerable interest in recent years owing to their benefit for applications ranging from precision spectroscopy in fundamental science,\cite{Newbury11} ultrafast ranging,\cite{Gaeta19} Tbit/s telecommunication systems,\cite{Gaeta19} and dual-comb broadband molecular spectoscopy\cite{Picque19} all the way to quantum signal and information processing.\cite{Kues19} Frequency comb generation in microresonators was first realized in 2007,\cite{DelHaye07} thereby reducing the footprint of such devices by orders of magnitude and making them even more appealing for applications. This first demonstration made use of the third-order $\chi^{(3)}$ nonlinear response, the so-called Kerr nonlinearities. This way of generating frequency combs in microresonators has thus been in the center of interest in recent years.\cite{Gaeta19, Chembo10, Kippenberg11, Hansson13, Kippenberg18,Pasquazi18} Here, the first comb lines are generated around the pump frequency $\nu_\mathrm{p}$ [Fig.\,\ref{fig1}a)] when the pump power threshold for $\chi^{(3)}$ hyperparametric oscillation is overcome.
When the power is raised further, subsequent four-wave mixing processes generate further comb lines, eventually building up a frequency comb.\\
There are, however, alternative ways of generating frequency combs in microresonators, based on cascaded $\chi^{(2)}$ nonlinear-optical processes. Since they are making use of the generally stronger $\chi^{(2)}$ nonlinearities, they potentially offer lower pump power thresholds and more flexibility regarding the pump frequency: in contrast with Kerr combs, that have to be pumped close to the zero-dispersion point, $\chi^{(2)}$ combs do not have such a limitation.\cite{Ricciardi15} Furthermore, because of the cascaded build-up of the $\chi^{(2)}$ processes involved, two combs are generated intrinsically [Fig.\,\ref{fig1}b-c)]. Generally, however, there is a large group velocity difference between the first- and second-harmonics leading to a walk-off, which was shown to be an important parameter for frequency comb generation.\cite{Leo16, Smirnov20} Also, phase-matching has to be assured for the $\chi^{(2)}$ processes. Sometimes, this can be achieved by employing different polarizations, i.e.\,birefringent phase-matching.\cite{Fuerst10shg} Usually, however, it requires the resonators to be radially poled.\cite{Beckmann11,Mohageg005}  \\
One approach is to start with $\chi^{(2)}$ optical parametric oscillation (OPO) as visualized in Fig.\,\ref{fig1}b). Subsequently, via second-harmonic generation, difference frequency generation and sum-frequency generation, combs are generated around both the pump frequency and half its frequency. As these processes are all thresholdless, the pump power threshold for OPO can be considered the frequency comb generation threshold. Such frequency combs were demonstrated experimentally in large bow-tie cavities\cite{Ulvila13, VU14, Mosca18} as well as very recently in microring resonators.\cite{Bruch2020} \\
Obviously, one can also pump at the lower frequency: the thresholdless second-harmonic generation is then followed by (internal) OPO, for which a pump power threshold needs to be overcome to generate the first comb lines [Fig.\,\ref{fig1}c)]. This initial step was demonstrated experimentally already over two decades ago in a cm-sized monolithic resonator.\cite{Schiller93} Subsequently, when raising the power, combs build up analogously to the previous case. This was demonstrated experimentally in bow-tie cavities,\cite{Ricciardi15} in waveguide resonators,\cite{Ikuta18} and, again very recently, in whispering gallery microresonators\cite{Szabados20, Hendry20} at pump powers as low as 2~mW.\cite{Szabados20} Although there is an analytical expression for the pump power threshold of internal OPO for Gaussian beams circulating in monolithic resonators with a few reflecting surfaces,\cite{Schiller93} such an expression is missing for whispering gallery microresonators.\\
In this contribution, we investigate the threshold for internal OPO, i.e.\,the frequency comb generation threshold in a whispering gallery resonator. Firstly, we model the field amplitudes of the interacting light fields in the cavity. This allows us to determine a simple analytic expression for the oscillation threshold. Then, we compare the oscillation thresholds for the three different frequency comb generation schemes sketched in Fig.\,\ref{fig1}. Finally, we experimentally determine the oscillation threshold and compare the observation with the prediction by our model.
\section{Theoretical considerations}
\begin{figure}
	\centering
	\includegraphics{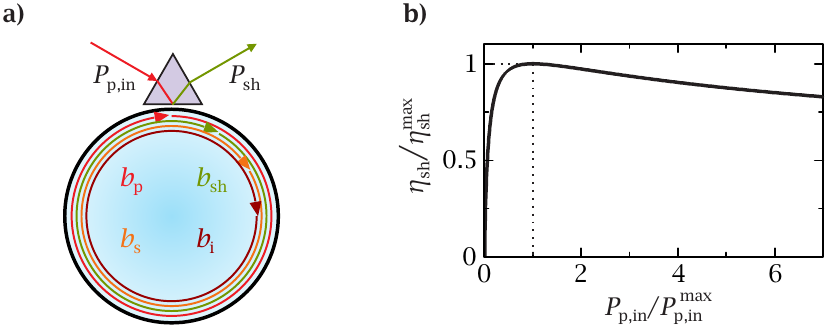}
	\caption{\textbf{a)} Schematic representation of the most important parameters needed for the theoretical description. Laser light with the power $P_\mathrm{p,in}$ is prism-coupled into a resonator. Inside the resonator, the pump light described by the complex internal amplitude $b_\mathrm{p}$ generates second-harmonic light ($b_\mathrm{sh}$). If this second-harmonic light overcomes the threshold for optical parametric oscillation (OPO), signal ($b_\mathrm{s}$) and idler ($b_\mathrm{i}$) light is generated. The power of the out-coupled second-harmonic light $P_\mathrm{sh}$ can be measured to yield the second-harmonic generation efficiency $\eta_\mathrm{sh}$, which is displayed in \textbf{b)}. Here, one can see that this efficiency increases with increasing incoupled pump power $P_\mathrm{p,in}$ until it reaches a maximum of $\eta_\mathrm{sh}^\mathrm{max}$ at a power $P_\mathrm{p,in}^\mathrm{max}$, slowly decreasing afterwards. }
	\label{figtheo}
\end{figure}

To derive a formula for the pump power threshold for frequency comb generation as visualized in Fig.\,\ref{fig1}c), we consider pump light with a certain power to be coupled into a whispering gallery microresonator. We assume phase-matching for second-harmonic generation to be fulfilled and the pump and second-harmonic light to be perfectly resonant. Furthermore, we only consider whispering gallery modes of one transverse type for the pump, signal, and idler light. To describe the situation inside the resonator, we use analytical expressions for the internal amplitudes $b_\mathrm{p,sh,s,i}$ referring to the pump, second-harmonic, signal, and idler light, respectively [Fig.\,\ref{figtheo}a)], with $\left|b_\mathrm{p,sh,s,i}\right|^{2}$ describing their intracavity powers,\cite{Sturman11} inserting these expressions into a notation introduced elsewhere.\cite{Breunig16}\\
With the incoupled pump power being $P_\mathrm{p,in}$, the second-harmonic generation efficiency can be calculated as 
\begin{equation}
\eta_\mathrm{sh}=\frac{P_\mathrm{sh}}{P_\mathrm{p,in}}=4\frac{r_\mathrm{sh}}{1+r_\mathrm{sh}}\frac{r_\mathrm{p}}{1+r_\mathrm{p}}\frac{X}{(1+X)^{2}},
\label{eq4}
\end{equation} 
where $P_\mathrm{sh}$ is the outcoupled second-harmonic power [Fig.\,\ref{figtheo}a)] and $r_\mathrm{p,sh}=\kappa_\mathrm{p,sh}^{2}/\alpha_\mathrm{p,sh}L$ are the ratios between coupling and internal loss for the pump and second-harmonic light respectively, with $\alpha_\mathrm{p,sh}$ being the absorption coefficient of the material the resonator is made of at the respective frequencies and $L$ the geometric circumference of said resonator.
The normalized intracavity pump power $X$ can be expressed as\cite{Sturman11}
\begin{equation}
X=\frac{8\nu_\mathrm{p}^{2}d^{2}L^{3}\mathcal{F}_\mathrm{sh}\mathcal{F}_\mathrm{p}}{c_{0}^{3}\epsilon_{0}n_\mathrm{p}^{2}n_\mathrm{sh}V}\left|b_\mathrm{p}\right|^{2},
\label{Bsi2}
\end{equation}
where $\nu_\mathrm{p}$ is the pump frequency, $d$ is the nonlinear-optical coefficient, $\mathcal{F}_\mathrm{p,sh}$ are the finesses at the pump and second-harmonic frequencies, respectively, $n_\mathrm{p,sh}$ are the respective refractive indices of the resonator material, $\epsilon_{0}$ is the vacuum permittivity, $c_{0}$ is the vacuum speed of light, and $V$ is the interaction volume. It can furthermore be expressed using the relation\cite{Breunig16}
\begin{equation}
X(1+X)^{2}=\frac{r_\mathrm{p}}{1+r_\mathrm{p}}\frac{P_\mathrm{p,in}}{P_{0}}
\label{eq2}
\end{equation} 
with the characteristic power $P_{0}$ being\cite{Breunig16}
\begin{equation}
P_{0}=\frac{\pi \nu_\mathrm{p}\epsilon_{0}}{8}\frac{n_\mathrm{p}^{4}n_\mathrm{sh}^{2}}{d^{2}}\frac{1}{Q_\mathrm{0p}^{2}Q_\mathrm{0sh}}V(1+r_\mathrm{p})^{2}(1+r_\mathrm{sh}),
\label{eq3}
\end{equation} 
where $Q_\mathrm{0p,0sh}$ are the intrinsic quality factors. Knowing all this, Eq.\,(\ref{eq2}) can be evaluated for different incoupled pump powers $P_\mathrm{p,in}$ to get $X$, which is then inserted into Eq.\,(\ref{eq4}). This yields the second-harmonic generation efficiency curve shown in Fig.\,\ref{figtheo}b), which has a maximum for $X=1$, i.e.\,at a pump power [Eq.\,(\ref{eq2})]
\begin{equation}
P_\mathrm{p,in}^\mathrm{max}=4P_{0}\frac{1+r_\mathrm{p}}{r_\mathrm{p}}.
\label{eqmax}
\end{equation}
We will now determine the incoupled pump power $P_\mathrm{th}^\mathrm{iOPO}$ needed to obtain a large enough internal second-harmonic power $\left|b_\mathrm{sh}\right|^{2}$ to overcome the pump threshold for internal OPO. The power of the second-harmonic wave inside the resonator is\cite{Sturman11}
\begin{equation}
\left|b_\mathrm{sh}\right|^{2}=\frac{8\nu_\mathrm{p}^{2}d^{2}L^{3}\mathcal{F}_\mathrm{sh}^{2}}{c_{0}^{3}\epsilon_{0}n_\mathrm{p}^{2}n_\mathrm{sh}V}\left|b_\mathrm{p}\right|^{4}.
\label{BSI1}
\end{equation}
If we now insert Eq.\,(\ref{Bsi2}) into Eq.\,(\ref{BSI1}), we obtain 
\begin{equation}
\left|b_\mathrm{sh}\right|^{2}=\frac{c_{0}^{3}\epsilon_{0}n_\mathrm{p}^{2}n_\mathrm{sh}V}{8\nu_\mathrm{p}^{2}d^{2}L^{3}\mathcal{F}_\mathrm{p}^{2}}X^{2}.
\label{Bsi3}
\end{equation}
Let us now turn to optical parametric oscillation. With our second-harmonic light acting as pump, for the internal amplitude of the generated signal light we find\cite{Sturman11}
\begin{equation}
b_\mathrm{s}=\frac{8\nu_\mathrm{s}\nu_\mathrm{i}d^{2}L^{3}\mathcal{F}_\mathrm{s}\mathcal{F}_\mathrm{i}}{c_{0}^{3}\epsilon_{0}n_\mathrm{s}n_\mathrm{i}n_\mathrm{sh}V}\left|b_\mathrm{sh}\right|^{2}b_\mathrm{s},
\label{Bsi4}
\end{equation} 
where the indices s and i stand for signal and idler light, respectively. As our comb generation relies on the ensuing signal and idler light to be nearly degenerate, we have $\nu_\mathrm{s}\approx\nu_\mathrm{i}\approx\nu_\mathrm{p}$ [Fig.\,\ref{fig1}c)], $\mathcal{F}_\mathrm{s}\approx \mathcal{F}_\mathrm{i}\approx\mathcal{F}_\mathrm{p}$, and $n_\mathrm{s}\approx n_\mathrm{i}\approx n_\mathrm{p}$. Inserting all this information into Eq.\,(\ref{Bsi4}) leads to
\begin{equation}
b_\mathrm{s}=\underbrace{\frac{8\nu_\mathrm{p}^{2}d^{2}L^{3}\mathcal{F}_\mathrm{p}^{2}}{c_{0}^{3}\epsilon_{0}n_\mathrm{p}^{2}n_\mathrm{sh}V}\left|b_\mathrm{sh}\right|^{2}}_{ T}b_\mathrm{s}.
\label{Bsi5}
\end{equation}
To get optical parametric oscillation, $T=1$ needs to be fulfilled. Thus, 
\begin{equation}
\left|b_\mathrm{sh}\right|_\mathrm{th}^{2}=\frac{c_{0}^{3}\epsilon_{0}n_\mathrm{p}^{2}n_\mathrm{sh}V}{8\nu_\mathrm{p}^{2}d^{2}L^{3}\mathcal{F}_\mathrm{p}^{2}}
\label{Bsi6}
\end{equation}
is the threshold for internally pumped OPO. By comparing Eqs.\,(\ref{Bsi3}) and (\ref{Bsi6}) we can immediately see that this is fulfilled when $X=1$. We can thus conclude that the threshold for internally pumped optical parametric oscillation, i.e.\,for frequency comb generation, coincides with a maximum in the second-harmonic generation efficiency $\eta_\mathrm{sh}$ [Eq.\,(\ref{eq4})] and is thus given by inserting Eq.\,(\ref{eq3}) into Eq.\,(\ref{eqmax}):
\begin{equation}
P_\mathrm{th}^\mathrm{iOPO}=P_\mathrm{p,in}^\mathrm{max}=\frac{\pi \nu_\mathrm{p}\epsilon_{0}}{2}\frac{n_\mathrm{p}^{4}n_\mathrm{sh}^{2}}{d^{2}}\frac{1}{Q_\mathrm{0p}^{2}Q_\mathrm{0sh}}V\frac{(1+r_\mathrm{p})^{3}(1+r_\mathrm{sh})}{r_\mathrm{p}}.
\label{eq5}
\end{equation}
Now that we have derived the pump power threshold for SHG-induced frequency comb generation, we can compare the thresholds of the three comb generation mechanisms shown in Fig.\,\ref{fig1}. \\
For frequency comb generation starting with OPO [Fig.\,\ref{fig1}b)], the pump power threshold is\cite{Breunig16}
\begin{equation}
P_\mathrm{th}^\mathrm{OPO}=\frac{\pi \nu_\mathrm{p}\epsilon_{0}}{16}\frac{n_\mathrm{p}^{2}n_\mathrm{s}^{2}n_\mathrm{i}^{2}}{d^{2}}\frac{1}{Q_\mathrm{0p}Q_\mathrm{0s}Q_\mathrm{0i}}V\frac{(1+r_\mathrm{p})^{2}(1+r_\mathrm{s})(1+r_\mathrm{i})}{r_\mathrm{p}},
\label{threshOPO}
\end{equation}
while the first comb lines in Kerr combs [Fig.\,\ref{fig1}a)] are generated when the pump power threshold for $\chi^{(3)}$ hyperparametric oscillation of
\begin{equation}
P_\mathrm{th}^{\mathrm{Kerr}}=1.54\frac{\pi\nu_\mathrm{p}}{4c_{0}}\frac{n_\mathrm{p}^{2}}{n_{2}}\frac{1}{Q_\mathrm{0p}^{2}}V\frac{(1+r_\mathrm{p})^{3}}{r_\mathrm{p}}
\label{threshkerr}
\end{equation}
is overcome,\cite{Ji17} see Appendix A. Here, $n_{2}$ is the nonlinear refractive index induced by the $\chi^{(3)}$ nonlinearity. If we assume $n_\mathrm{p,sh,s,i}=2$, $V=10^{-12}$~m$^{3}$, $Q_\mathrm{0p,0sh,0s,0i}=10^{7}$, and $r_\mathrm{p,sh,s,i}=1$, for a pump wavelength of $\nu_\mathrm{p}=300$~THz and a nonlinear refractive index of $n_{2}=10^{-19}$~m$^{2}$/W, the pump threshold for Kerr comb generation [Fig.\,\ref{fig1}a)] is $P_\mathrm{th}^\mathrm{Kerr}=3.9$~W [Eq.\,(\ref{threshkerr})]. To get frequency combs around this frequency and double the frequency, assuming $d=10^{-12}$~m/V and $\nu_\mathrm{p}=600$~THz, the pump threshold for a $\chi^{(2)}$-comb starting with an OPO process [Fig.\,\ref{fig1}b)] is $P_\mathrm{th}^\mathrm{OPO}=1.1$~mW according to Eq.\,(\ref{threshOPO}), while the same value for a comb starting with second-harmonic generation [Fig.\,\ref{fig1}c)], i.e.\,pumped at $\nu_\mathrm{p}=300$~THz, is $P_\mathrm{th}^\mathrm{iOPO}=4.3$~mW using Eq.\,(\ref{eq5}). This furthermore underlines the potential of $\chi^{(2)}$ combs coming with very low pump power thresholds; it should be noted, however, that highly optimized Kerr combs, where the interaction volume $V$ was on the 1000-\textmu m$^{3}$-scale and below, were already demonstrated with sub-mW pump thresholds.\cite{Ji17, ZhangDel2019,Chang20}
\section{Experimental methods}
\begin{figure*}
	\centering
	\includegraphics{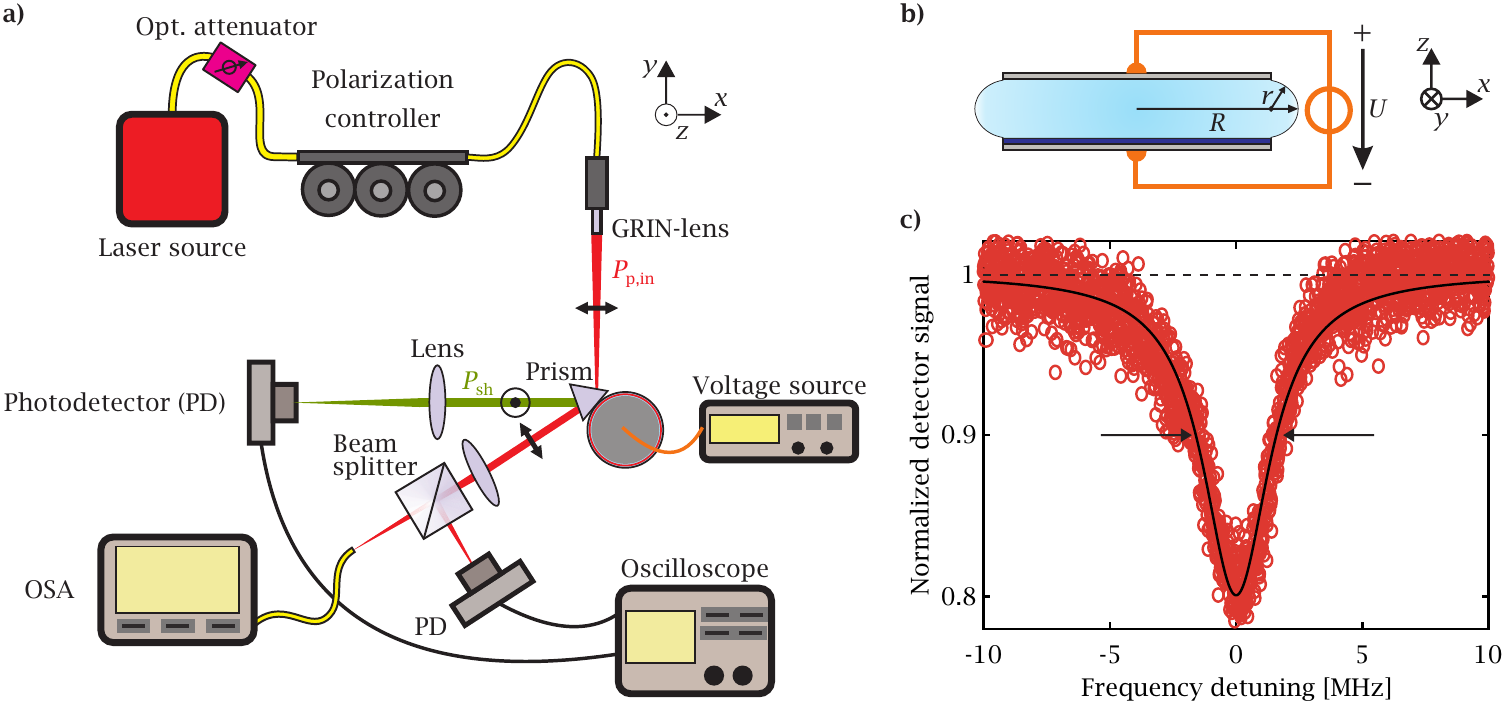}
	\caption{\textbf{a)} Sketch of the experimental setup. Light from a continuous-wave laser source is prism-coupled into a microresonator, which is kept at temperatures $T\approx 70~^{\circ}$C to fulfil birefringent phase-matching for second-harmonic generation. Light around the pump frequency and the second-harmonic frequency are spatially separated because of the large birefringence of the rutile prism we employ. The light around the pump frequency passes a beam splitter: this way, we can monitor the transmission spectrum with a photodetector as well as the generated spectra using an optical spectrum analyzer (OSA). The second-harmonic light is also focused on a photodetector so we can measure its power. \textbf{b)} Side-view of the whispering gallery resonator with major radius $R=1$~mm and minor radius $r=380$~\textmu m. \textbf{c)} In critical coupling, we can determine the cavity resonance linewidth to be 3.2~MHz and the maximum coupling efficiency to be 20~\%.}
	\label{Fig2}
\end{figure*}

To validate the model introduced in the previous section, we couple laser light into a whispering gallery resonator and measure the generated second-harmonic power $P_\mathrm{sh}$ as shown in Fig.\,\ref{figtheo}a) to determine the second-harmonic generation efficiency $\eta_\mathrm{sh}$ [Eq.\,(\ref{eq4})]. By changing the incoupled pump power, we can thus determine an efficiency curve as shown in Fig.\,\ref{figtheo}b). Simultaneously, we monitor the generated spectrum around the pump frequency to observe the onset of internally pumped optical parametric oscillation, i.e.\,the frequency comb generation threshold $P_\mathrm{th}^\mathrm{iOPO}$.\\
We carry out this experiment with the experimental setup sketched in Fig.\,\ref{Fig2}a). 
Its central part is a whispering gallery resonator made of 5\% MgO-doped $z$-cut congruent lithium niobate (CLN) with a geometry as shown in Fig.\,\ref{Fig2}b). To manufacture such a resonator, we start with a 300-\textmu m-thick CLN wafer with chromium deposited on the $+z$-side. From this wafer, we cut out a cylinder using a femtosecond laser source emitting at 388~nm wavelength with 2~kHz repetition rate and 300~mW average output power. This cylinder is then glued with its $-z$-side to a metal post for easier handling. Subsequently, the same femtosecond laser source is employed to give the resonator the desired geometry of a major radius $R=1$~mm and a minor radius $r=380$~\textmu m (Fig.\,\ref{Fig2}b)). To achieve an optical-grade surface quality, the resonator is eventually polished with a diamond slurry.\\
To couple light into this resonator, we employ a frequency-tunable fiber-coupled continuous-wave laser source emitting around 1064~nm wavelength (NKT Koheras Basik, 20~kHz linewidth) passing an optical attenuator (Thorlabs V1000A) allowing to set the power and a polarization controller to set the polarization of the pump light. A gradient-index (GRIN) lens focuses the light onto a rutile prism, which is in close proximity to the whispering gallery resonator rim; this way, light can be coupled into the resonator. Firstly, we characterize the microresonator at very low pump powers in order to avoid thermal and nonlinear-optical effects on the linewidth. This way, we can determine the quality factor and the coupling efficiency. In the following, we set the distance between prism and resonator rim such that we are in slight undercoupling, i.e.\,$r_\mathrm{p}<1$ as from our model we expect the pump thresholds to be lower here [Eq.\,(\ref{eq5})].\\
To fulfil the phase-matching condition for second-harmonic generation, we heat the resonator to $T\approx 70~^{\circ}$C and use the birefringence of CLN: the pump light is polarized in the $x$-$y$-plane of the crystal, thus experiencing the ordinary (o-) refractive index of CLN. This leads to the generation of extraordinarily (e-) polarized second-harmonic light.\cite{Fuerst10shg} The generated signal and idler light is again o-polarized; eventually, the comb around the pump frequency is o-polarized, while the one around the second-harmonic is e-polarized. The large birefringence of the rutile prism automatically separates the light around the pump and the second-harmonic frequencies spatially.\cite{Szabados20} To fulfil the prerequisite of the interacting pump and second-harmonic waves being perfectly resonant, we connect the top electrode of the resonator and the metal post below to a voltage source [Fig.\,\ref{Fig2}b)]. Then, by applying a bias voltage and tuning the laser to stay zero-detuned from the cavity resonance, we can maximize the second-harmonic power, thus making sure the pump and second-harmonic waves are perfectly resonant.\cite{Fuerst10shg} The outcoupled o-polarized light around the pump frequency finally passes a beam splitter: part of it hits a calibrated silicon photodetector so we can measure the normalized transmission spectrum of the pump light as well as the pump power, while a larger part enters an optical spectrum analyzer (Yokogawa AQ6370D) so we can monitor the generated spectra around the pump frequency, thus allowing us to determine the onset of OPO, i.e.\,the frequency comb generation threshold. The second-harmonic light is also directed to a calibrated silicon photodetector, so we can measure the second-harmonic power $P_\mathrm{sh}$.      
\section{Results and discussion}
We find the maximum coupling efficiency to be 20~\% and the intrinsic quality factor to be $Q_\mathrm{0p}=1.7\times10^{8}$ [Fig.\,\ref{Fig2}c)], which is close to the absorption limit of CLN.\cite{Leidinger15}  
\begin{figure}
    \centering
    \includegraphics{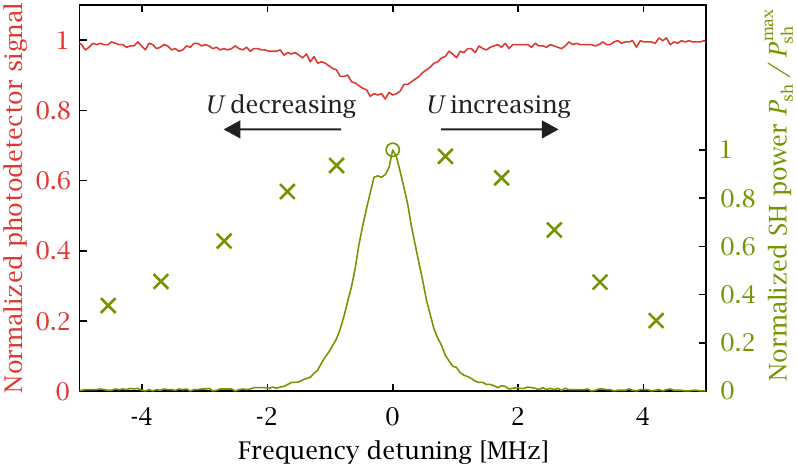}
    \caption{When scanning the pump laser over a cavity resonance (red), second-harmonic light is generated (solid green curve). By applying a bias voltage $U$ to the resonator, the pump as well as cavity resonances at the second-harmonic frequency are shifted relative to each other. The green crosses mark the peak of the power of the second-harmonic light for different applied voltages, with $\Delta U=50$~mV for two adjacent measurements, while the circle stands for the maximum achievable second-harmonic power. Using this approach, we can maximize the second-harmonic power for each pump power $P_\mathrm{p,in}$. }
    \label{Fig3}
\end{figure}
As mentioned in the previous section, we then determine the second-harmonic generation efficiency. A typical measurement showing both the normalized transmission of the pump light as well as the generated second-harmonic light is shown in Fig.\,\ref{Fig3}. By applying external bias voltages $U$ of up to a few hundred mV, we shift the frequency of the pump resonance as well as of the second-harmonic resonance; by minimizing the detuning of the two waves, we can maximize the second-harmonic power $P_\mathrm{sh}$ as previously explained. We repeat this process for a number of different incoupled pump powers $P_\mathrm{p,in}$ by using the optical attenuator in our setup [Fig.\,\ref{Fig2}a)]. The resulting second-harmonic generation efficiency curve is visualized in Fig.\,\ref{Fig4}a).
\begin{figure*}
	\centering
	\includegraphics{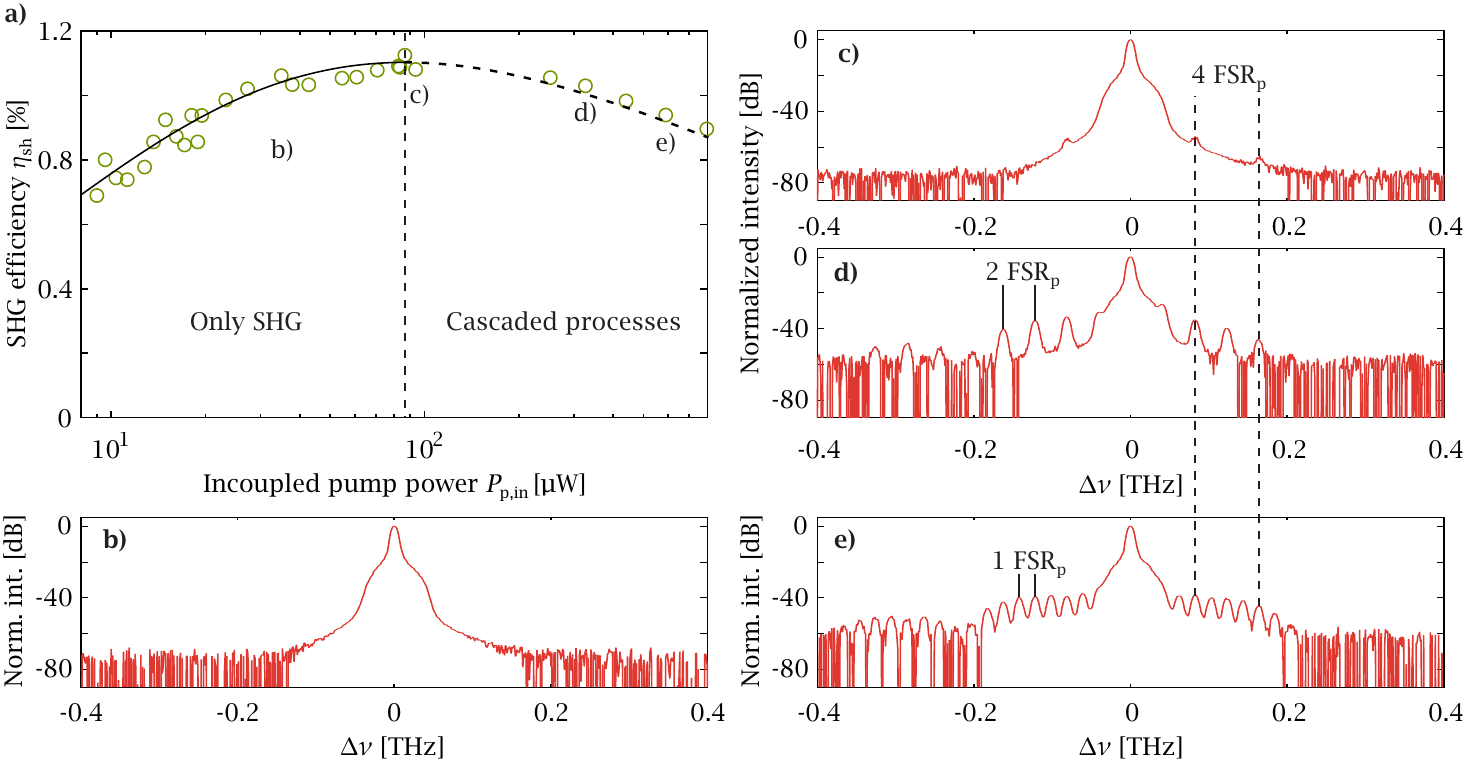}
	\caption{\textbf{a)} The second-harmonic generation efficiency increases with increasing incoupled pump powers $P_\mathrm{p,in}$ until it reaches a maximum at 85~\textmu W. The solid black line shows the theoretical curve [Eq.\,(\ref{eq4})] describing this behavior. For pump powers below 85~\textmu W, we only observe the pump laser on the optical spectrum analyzer as shown in \textbf{b)}. At the maximum of $\eta_\mathrm{sh}$, however, internally pumped optical parametric oscillation sets in, see \textbf{c)} - this can be considered the first step of frequency comb generation. Further increasing the pump power leads to more comb lines becoming possible to observe, see \textbf{d)}, with the initial separation between individual lines of 4~FSR$_\mathrm{p}$ reducing to 2~FSR$_\mathrm{p}$ at 330~\textmu W. At even higher pump powers of 590~\textmu W, finally, we start observing a frequency comb with only 1~FSR$_\mathrm{p}$ between individual components; this is visualized in \textbf{e)}. For pump powers above the comb generation threshold, the theoretical curve is shown as a dashed line, as here the prerequisite of the pump and second-harmonic light being perfectly resonant is not necessarily fulfilled. }
	\label{Fig4}
\end{figure*}
As one can clearly see, $\eta_\mathrm{sh}$ grows with increasing pump power until it reaches its maximum at $P_\mathrm{p,in}\approx 85$~\textmu W. When looking at the spectrum around the pump frequency, one can see that while for $P_\mathrm{p,in}<85$~\textmu W we only observe the pump field [Fig.\,\ref{Fig4}b)], the peak in second-harmonic generation efficiency is accompanied by the onset of optical parametric oscillation [Fig.\,\ref{Fig4}c)], just as predicted by the model above. We can even already observe further equally spaced sidebands, which comes as no surprise as the OPO threshold is the only one that needs to be overcome for comb generation with the other processes being thesholdless. Thus, $P_\mathrm{th}^{\mathrm{iOPO}}=85$~\textmu W can be considered the pump threshold for frequency comb generation in our system. To the best of our knowledge, this is the lowest pump threshold reported for $\chi^{(2)}$-based frequency combs so far. Analogously to previously published work, the comb spacing locks to multiples of the free spectral range (FSR) at the pump frequency, FSR$_\mathrm{p}=20.8$~GHz, despite a large FSR-offset between the pump and second-harmonic waves of 1.3~GHz,\cite{Szabados20} with the initial spacing being 4~FSR$_\mathrm{p}$. When further raising the pump power, at $P_\mathrm{p,in}\approx 330$~\textmu W, we can see the generated comb lines being separated by 2~FSR$_\mathrm{p}$ [Fig.\,\ref{Fig4}d)], finally ending up with the desired 1~FSR$_\mathrm{p}$ comb line spacing at $P_\mathrm{p,in}\approx590$~\textmu W [Fig.\,\ref{Fig4}e)]. This transition from 4~FSR$_\mathrm{p}$ to 1~FSR$_\mathrm{p}$ is yet unexplained and needs to be investigated further.\\
As already stated, the OPO onset in our measurements coincides nicely with the maximum in second-harmonic generation efficiency, just as predicted by the model developed above. In the following, we compare the experimental results with our prediction quantitatively. With our pump frequency being $\nu_\mathrm{p}=281.749$~THz, at temperatures $T\approx70~^{\circ}$C we can determine the bulk refractive indices of CLN to be $n_\mathrm{p}=2.2291$, $n_\mathrm{sh}=2.2263$;\cite{Umemura16} for a resonator of the size used here this is a good approximation for the effective refractive indices of the modes. The nonlinear-optical coefficient for second-harmonic generation is $d=4.7$~pm/V.\cite{Eckardt90} Furthermore, we determined $Q_\mathrm{0p}=1.7\times10^{8}$: if we now employ Eq.\,(\ref{eq4}) on our measured second-harmonic generation efficiencies, we find a very good match between the experimental data and the model for $Q_\mathrm{0sh}=1.7\times10^{7}$, $r_\mathrm{p}=0.5$, $r_\mathrm{sh}=0.034$, and $V=280\times10^{-12}$~m$^{3}$. As we know from literature that for the frequencies involved it is $\alpha_\mathrm{sh}\approx 10\alpha_\mathrm{p}$,\cite{Leidinger15}, $Q_\mathrm{0sh}=0.1Q_\mathrm{0p}$ is a reasonable assumption. As we carried out our measurements in the undercoupling regime, $r_\mathrm{p}<1$ matches well with our expectations. Furthermore, $r_\mathrm{sh}<0.1r_\mathrm{p}$ has to be the case as $\kappa_\mathrm{sh}<\kappa_\mathrm{p}$ for the wavelengths involved and $r_\mathrm{p,sh}=\kappa_\mathrm{p,sh}^{2}/\alpha_\mathrm{p,sh}L$ as mentioned in the introduction.\cite{Breunig16} Taking the size of our resonator into account, the interaction volume $V$ can also be considered reasonable. Thus, the model introduced above describes our experimental data very well up until $\eta_\mathrm{sh}^{\mathrm{max}}$ at $P_\mathrm{th}^{\mathrm{iOPO}}=P_\mathrm{p,in}^\mathrm{max}$. For higher pump powers $P_\mathrm{p,in}>P_\mathrm{th}^{\mathrm{iOPO}}$, we cannot be certain about the pump and second-harmonic waves being perfectly resonant as the system is optimized to yield the best-possible comb signals. 
\section{Conclusion}
We demonstrate $\chi^{(2)}$ frequency comb generation in a lithium niobate whispering gallery microresonator at very low pump power thresholds of 85~\textmu W, observing the onset of internal optical parametric oscillation, i.e.\,the comb threshold, to coincide with the maximum in the second-harmonic generation efficiency. Building on previously published modeling of these processes,\cite{Sturman11, Breunig16} we show that this is not just a coincidence, but a behavior that is also expected from a theoretical point of view. With the newly found analytical formula for the pump power threshold [Eq.\,\ref{eq5}], we can estimate that the pump power threshold might potentially be lowered by five orders of magnitude by reducing the interaction volume to roughly 1000~\textmu m$^{3}$, a value that is achievable using chip-integrated lithium niobate microresonators; thus far, however, the achievable quality factors in such resonators are about an order of magnitude lower than those of the hand-polished, larger resonators we use,\cite{ZhangHarv17} which affects the pump power threshold as $P_\mathrm{th}^\mathrm{iOPO}\propto Q_\mathrm{0p}^{-2}Q_\mathrm{0sh}^{-1}$ [Eq.\,(\ref{eq5})]. Even with current state of the art values, however, the pump power thresholds should be in the sub-\textmu W range already, making such chip-integrated $\chi^{(2)}$ frequency comb sources very appealing for applications. \\
While the frequency comb shown here relies on modulation instabilities analogously to previous experiments,\cite{Szabados20,Mosca18} recently, for the first time, solitonic behavior was observed in a chip-integrated $\chi^{(2)}$ comb on an aluminum nitride platform at one of the two wavelengths involved.\cite{Bruch2020} Several theoretical studies argue that it is possible to generate solitonic combs with the scheme shown in our paper if the phase-matching is accompanied by a reduced or vanishing FSR-offset between the pump and second-harmonic.\cite{Villois19a,Villois19b, Smirnov20} Such matching FSRs spectrally separated by an octave can be found in the present sample: in that case, however, one needs to radially pole the resonator, i.e.\,a quasi-phase-matching (QPM) structure needs to be employed.\cite{Breunig16} The use of a QPM-structure would allow for great flexibility regarding the choice of center wavelengths and polarization of the generated frequency combs. Thus, future work will focus on realizing solitonic $\chi^{(2)}$ frequency combs by introducing QPM-structures to our microresonators and realizing $\chi^{(2)}$ microcomb sources on a batch-compatible chip-integrated platform. 
\begin{acknowledgments}
This work was supported by the Fraunhofer and Max Planck Cooperation programme. The authors would like to thank Karsten Buse for his advice on the project and his feedback on the manuscript as well as Yannick Minet for providing the microresonator. 
\end{acknowledgments}
\appendix
\section{Kerr comb threshold}
The formula given in literature looks the following:\cite{Ji17} 
\begin{equation}
P_\mathrm{th}^\mathrm{Kerr}=1.54\frac{\pi}{4}\frac{n_\mathrm{p}^{2}}{\lambda_\mathrm{p}n_{2}}\frac{Q_\mathrm{cp}}{Q_\mathrm{lp}^{3}}V.
\label{ThreshKerrOrig}
\end{equation}
Here, $\lambda_\mathrm{p}=c_{0}/\nu_\mathrm{p}$ is the pump wavelength and $Q_\mathrm{cp}$ ($Q_\mathrm{lp}$) is the coupled (loaded) quality factor. We can express the loaded quality factor as
\begin{equation}
Q_\mathrm{lp}^{-1}=Q_\mathrm{0p}^{-1}+Q_\mathrm{cp}^{-1}
\end{equation}
with $Q_\mathrm{0p}=2\pi n_\mathrm{p}/(\lambda_\mathrm{p}\alpha_\mathrm{p})$ and $Q_\mathrm{cp}=2\pi n_\mathrm{p} L/(\lambda_\mathrm{p}\left|\kappa\right|^{2})$.\cite{Sturman11} With the coupling parameter $r_\mathrm{p}=\left|\kappa\right|^{2}/(\alpha_\mathrm{p} L)$, we can thus express the coupling quality factor as 
\begin{equation}
Q_\mathrm{cp}=\frac{2\pi n_\mathrm{p} L}{\lambda_\mathrm{p}\left|\kappa\right|^{2}}=Q_\mathrm{0p}\frac{1}{r_\mathrm{p}}
\label{coupledQ}
\end{equation}
and the loaded quality factor as 
\begin{equation}
Q_\mathrm{lp}=\frac{2\pi n_\mathrm{p} L}{\lambda_\mathrm{p}(\alpha_\mathrm{p} L+\left|\kappa\right|^{2})}=Q_\mathrm{0p}\frac{1}{1+r_\mathrm{p}}.
\label{loadedQ}
\end{equation}
Inserting Eqs.\,(\ref{coupledQ},\ref{loadedQ}) into Eq.\,(\ref{ThreshKerrOrig}), we thus end up with
\begin{equation}
P_\mathrm{th}^{\mathrm{Kerr}}=1.54\frac{\pi\nu_\mathrm{p}}{4c_{0}}\frac{n_\mathrm{p}^{2}}{n_{2}}\frac{1}{Q_\mathrm{0p}^{2}}V\frac{(1+r_\mathrm{p})^{3}}{r_\mathrm{p}},
\end{equation}
exactly as stated in Eq.\,(\ref{threshkerr}).
\bibliography{paperbib}

\end{document}